# A highly sensitive piezoresistive sensor based on MXene and polyvinyl butyral with a wide detection limit and low power consumption

*Ruzhan Qin*[†], *Mingjun Hu*[‡], *Xin Li*[†], *Li Yan*[‡], *Chuanguang Wu*[‡], *Jinzhang Liu*[‡], Haibin Gao[⊥], *Guangcun Shan*[†]*, *and Wei Huang*[§]*

[†]School of Instrumentation Science and Opto-electronics Engineering, Beihang University, Beijing 100191, P. R. China

[‡]School of Materials Science and Engineering, Beihang University, Beijing 100191, P. R. China

[⊥] Institute of Experimental Physics, Saarland University, 66123 Saarbrücken, Germany

[§]Shaanxi Institute of Flexible Electronics, Northwestern Polytechnical University, Xi'an 710072, P. R.

China

* Corresponding authors:
gcshan@buaa.edu.cn; iamwhuang@nwpu.edu.cn

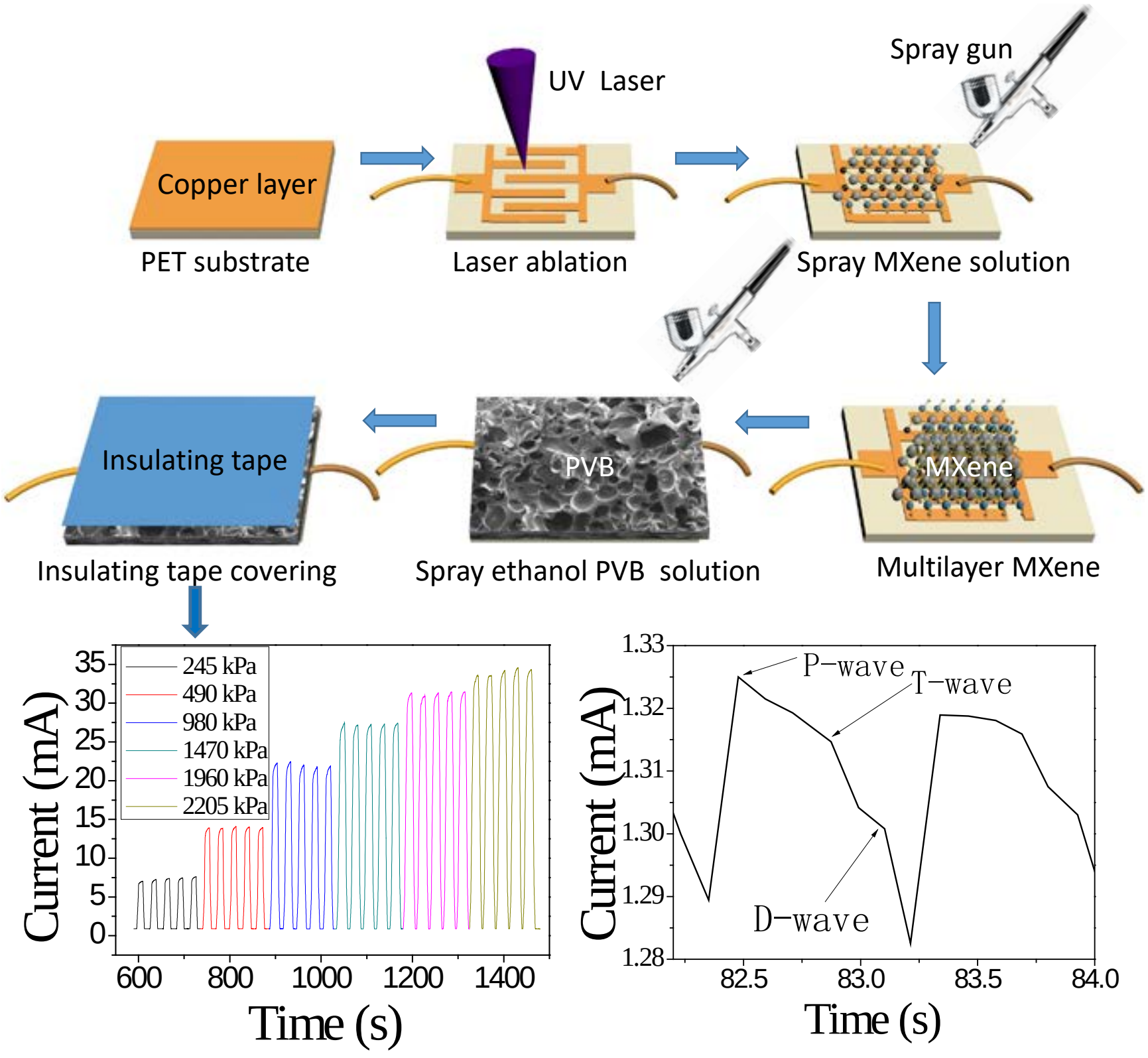

**ABSTRACT:**

As a new class of two-dimensional transition-metal carbide and carbonitride, MXenes have been widely used in the energy storage, sensor, catalysis, electromagnetic shielding, and other field. Currently it is challenging to fabricate highly sensitive MXene-based flexible sensors with low power consumptions and wide detection limits. In this work, taking advantage of the porous structure of polyvinyl butyral (PVB) and MXene, we fabricate a highly flexible and highly sensitive piezoresistive sensor that is low power consuming and exhibits a wide detection limit. The fabricated flexible sensor is highly sensitive (Gauge factor: ~202.2) and exhibits a response time of ~110 ms. The MXene/PVB-based sensor has a wide detection range (49 Pa to ~2.205MPa), which can detect a low voltages of 0.1 mV, consumes little power (~$3.6 \times 10^{-10}$ W), is very mechanically stable (almost no attenuation over 10,000 maximum-pressure cycles), and is simple to prepare, among other features. The MXene/PVB-based flexible sensor can detect subtle bending and release during human activities, arterial pulses, and voice signal. The as-prepared MXene/PVB-based flexible sensor is reliable and endurable, which is potentially suitable as a high-pressure, highly sensitive, and energy-efficient piezoresistive sensor with a wide detection range.

## INTRODUCTION

MXene is a graphene-like two-dimensional layered material composed of transition-metal carbides/nitrides and carbonitrides, and is prepared from a MAX phase, a family of more than 70 layered three-dimensionally structured ternary carbides/nitrides and carbonitrides. MXene, with surfaces that contain rich active groups (-F, -O, and -OH groups), is prepared by selective etching the "A" from MAX with hydrofluoric acid or a LiF/HCl solution. The general chemical formula of MXene is $M_{n+1}X_nT_x$, where M is the transition metal, X is C and/or N, and T represents a surface reactive end group (n = 1, 2, or 3). Because of its similar structure to that of graphene, these materials are referred to as "MXene" [1-2]. $Ti_3AlC_2$ is a well-studied MAX-phase material from which the Al is selectively etched to form $Ti_3C_2Tx$ nanosheets. Since the first successful preparation of a two-dimensional transition-metal carbon/nitride by stripping in 2011[1], MXene have attracted significant levels of attention in many fields due to their superior properties and application prospects.

The excellent electronic, optical, mechanical, and thermal properties of MXene have led to their extensive explorations in supercapacitors[3-5], electromagnetic interference shielding[6], photodetectors[7-8], field effect transistors[9], sensors, including gas[10-11], bio[12-15], humidity[16-17], electrochemical[18-19], strain[20-23], and pressure[24-30], detection applications[31-32], electrocatalysis[33-34], nanogenerators[35] and so on. In particular, piezoresistive sensors that use the greatly altered interlayer distance of MXene under external pressure, which is a basic MXene characteristics, have been reported[24]. However, in order to improve device performance, most sensors have to be combined with other materials, such as sponges, porous materials, and polymers [25-30]. In fact, pressure sensors reported so far commonly include basic pressure, strain, shear, and vibration types, and their combinations. The changes in sensor resistance are mainly due to changes in sensor geometry, the semiconductor band-gap, in the contact resistance between two materials, and in the particle spacings in composite materials[36], with sensing mechanisms mainly based on pressure, capacitance, piezoelectric, and optical effects[24,25,30,36]. A piezoresistive sensor is a type of sensor that converts the change in material resistance stimulated by external pressure into an electrical output with the advantages of a simple structure, low preparation costs, and ease of signal acquisition and data output, which have important applications in electronic display equipment, electronic skin, and wearable medical monitoring instruments.

Sensitive materials reported for use in flexible sensors include nanoparticles[20], nanowires[37], carbon nanotubes[5], graphene[6,30], and organic materials[25,27,28], and these sensitive materials are prepared by spraying, vacuum osmosis, printing, and dipping, among others. However, these processes are often relatively complicated, and the sensing materials tend to become detached under external force. Therefore, simple coating methods can lead to sensor instability[24]. In addition, due to the unique internal atomic structures of graphene and carbon nanotubes, their very high Young's moduli resist the movements of their constituent atoms; consequently, it is difficult to further improve the sensitivities of the corresponding sensors[38]. Owing to its greatly changed interlayer distance of the lamellar structure, MXene is highly sensitive and flexible under an external pressure, which satisfies practical requirements. Flexible substrates, including polyimide (PI), polyethersulfone (PES), polyethylene naphthalate (PEN), and polyethylene terephthalate (PET) are often selected for sensing applications[39]. Flexible sensors are able to stretch and bend due to the flexibility of the substrate as well as that of the packaging materials of the devices, which are usually fabricated from PDMS[9,20,22,24,25,27-29,37,39].

Flexible wearable sensors that can detect a wide range of physical activities, including large

hand[20,21,24,26], arm[24], and leg[20,23,24] movements, finger bending[20,21,22,24,25,27-29], as well as smaller breathing[20,27], blowing[20], and swallowing[24-26] movements, vocal-muscle vibrations[20,28-30], pulses [20,21,23,25,26,28-30], and eye pressure[20,24], have been rapidly evolving.

However, simultaneously realizing extremely high sensitivity, wide detection limits, and high strain coefficients, as well as low power-consumption, good mechanical stability, and a simple preparation process, remain challenges. As far as we know, there have been no reports on combining MXene with polyvinyl butanal (PVB) to fabricate flexible wearable pressure sensors. In recent years, PVB has been used to prepare a variety of sensor devices that benefit from the nano-porous structure and excellent mechanical stability of the PVB polymer. It should be noted that PVB is a type of adhesive material that can improve the sensitivity and detection limit of an electronic device, while also exhibiting corrosion-resistance properties[40-45]. Herein, we take advantages of MXene and PVB to fabricate a piezoresistive sensor that exhibits a wide detection limit (~49Pa to ~2.205 MPa), high sensitivity (Gauge factor (GF) ~202.2; response time ~110 ms), low detection voltage (0.1 mV), consuming little power (~$3.6 \times 10^{-10}$ W) via a simple preparation process, which is highly mechanically stable (strain ~1.25%; over 10,000 maximum pressure cycles). The state-of-the-art of our MXene/PVB-based flexible sensor here can detect subtle bending and release during human activities, arterial pulses, and other weak pressures.

## RESULTS AND DISCUSSION

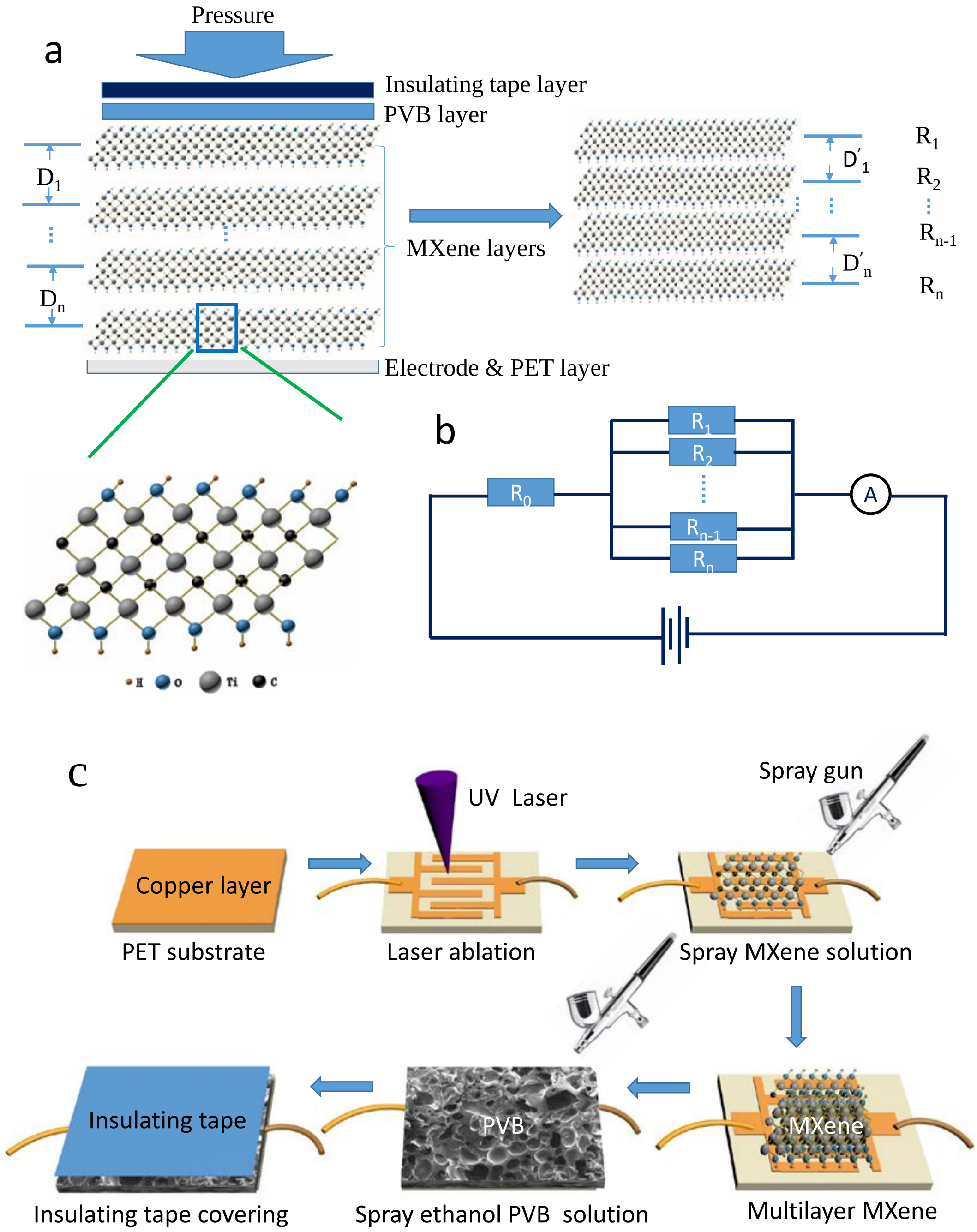

**Figure 1.** The structure and mechanism of a piezoresistive sensor based on MXene/PVB. (a) The MXene/PVB-based sensor consists of four parts: the bottom PET layer with the interdigital fingers electrode, the middle MXene layers, the PVB layer, and the upper insulating tape layer. The gaps between adjacent MXene layers are compressed under external pressure, but the distances between lattices of the same MXene layer do not changed under external pressure. (b) Equivalent circuit diagram of a piezoresistive sensor based on MXene/PVB that shows that the total resistance decreases with decreasing inter-layer distance. The parallel total resistance decreases as the distance between layers decreases; hence, the total resistance also decreases. (c) Showing the process for the preparation of the MXene/PVB-based sensor.

Figure 1 shows the structure and the mechanism of the MXene/PVB-based piezoresistive sensor.

The sensor contains four main parts, namely the bottom PET layer with an interdigital finger electrode, the middle MXene layers, the PVB layer, and the upper insulating tape layer. As shown in Figure 1(a), the distance between two neighbouring MXene layers decreases under external pressure (such as $D_1 \rightarrow D'_1, \cdots, D_n \rightarrow D'_n$), which results a decrease in the total resistance of the MXene layers. The resistance of the MXene layers decreases mainly because the MXene layers are equivalent to a number (n) of parallel resistors. The MXene layers maintain a stable resistance value, which is the initial value in the absence of external pressure. Figure 1(b) shows the equivalent circuit diagram, where the resistance of the electrode circuit and the wires of the sensor are assumed to be $R_0$. The resistance of the sensor MXene layers remains constant in the absence of a loading pressure, but when subjected to an external pressure, the resistance of these layers is assumed to be variable ($R_C$). Here $R_C = R_1, R_2, \cdots, R_{n-1}$, and $R_n$ parallel resistances; hence the expression for the total resistance ($R_{total}$) of the sensor becomes:

$$R_{total} = R_0 + R_C \tag{1}$$

with $R_C$ determined from the following formula

$$\frac{1}{R_C} = \frac{1}{R_1} + \frac{1}{R_2} + \cdots + \frac{1}{R_n} \tag{2}$$

where, $R_{total}$, $R_0$, And $R_C$ are the total resistance of the sensor, the resistance of the electrode circuit and electrical wires, and the variable resistance of the MXene, respectively. Changes in external pressure can therefore be monitored by the changes in resistance.

Copper clad flexible PET film, with a ~200-nm-thick copper layer and a ~200-μm-thick PET layer was purchased commercially. As shown in Figure 1(c), the copper film on the PET was selectively removed through laser ablation to create the required conduction pathway. The interdigital electrode and display on the flexible PET substrate was prepared using a laser ablation strategy[46]. In order to ensure sensor conduction, conductive silver paste or other conducting strips can be used as the lead to the interdigital finger electrode. The size of the interdigital electrode can be designed according to the specific requirements of the sensor. In this work, a 1.8 cm × 1.5 cm interdigital finger electrode was used. The distance between two adjacent electrode fingers is about 540 ± 20 μm, while a single finger is 900 ± 20 μm long and 1000 ± 20 μm wide, as shown in Figure S1. Compared with other interdigital electrode manufacturing technologies, the laser ablation strategy is simple, convenient, and non-toxic. The specific preparation process is shown in Figure 1(c). The MXene layers are then sprayed onto the interdigital finger circuit on the soft PET with a spray gun. MXene thickness is determined by the required resistance range of the sensor. PVB dissolved in ethanol is then sprayed onto the MXene layers, and ordinary insulating tape is then used as a protective layer on the PVB layer, to ensure that the MXene does not become detached under external mechanical stress. Polydimethylsiloxane (PDMS) can be used as final packaging in order to ensure the long-term use of the sensor.

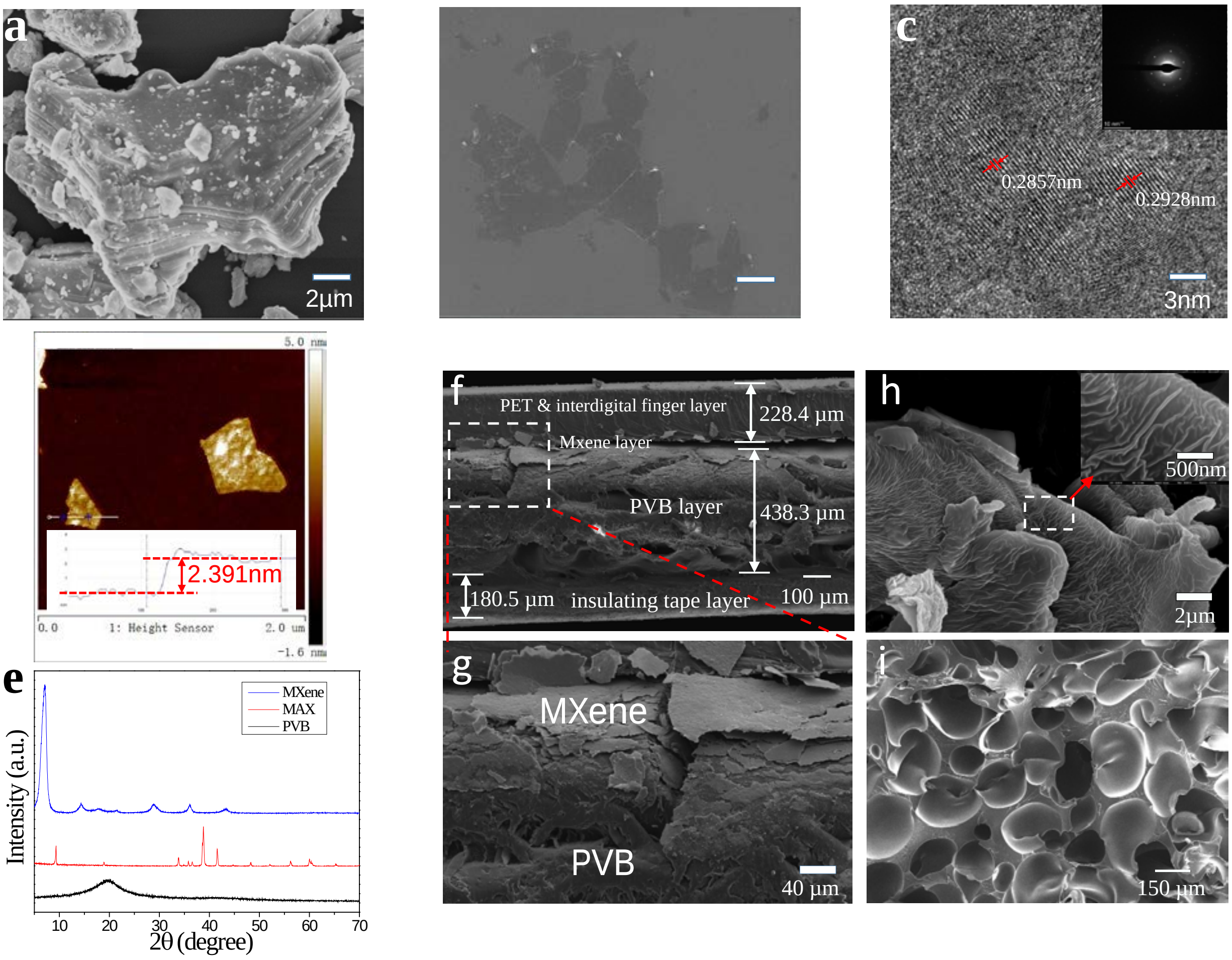

**Figure 2.** Characterising the material of MXene/PVB-based sensor. (a) SEM image of MAX ($Ti_3AlC_2$) showing that layers have not peeled off. (b) SEM image of MXene ($Ti_3C_2T_x$). (c) Planar view HRTEM image of MXene and its corresponding diffraction pattern in the inset, indicating that the distance between MXene layers is ~0.29 nm. (d) Atomic force microscopy (AFM) image of MXene which shows that the MXene nanosheets are ~2.391 nm thick. (e) The X-ray diffraction (XRD) patterns of MAX, MXene, and PVB. (f) Cross-sectional SEM image of the MXene/PVB-based sensor, containing the PET+ interdigital-finger-electrode layer, the MXene layer, the PVB layer, and insulating tape layer. (g) Locally enlarged SEM image corresponding to the indicated region in (f). (h) SEM image of the folded structure of PVB after PVB was spraying onto the MXene layer. (i) SEM image of the porous structure of the raw PVB material.

MXene ($Ti_3C_2T_x$) nanosheets were fabricated by selectively etching the Al layer from $Ti_3AlC_2$ (Figure 2(a)) using a mixed solution of LiF and hydrochloric acid as the etchant, and the few layers structure of MXene (Figure 2(b)) is confirmed by scanning electron microscopy (SEM). Figure 2(c) shows the HRTEM image of MXene, from which the distance between MXene layers is determined to be ~0.29 nm, with the corresponding diffraction pattern shown in the inset which showing the hexagonal feature. Atomic force microscopy (AFM) reveals that the $Ti_3C_2T_x$ nanosheets are 2.391 nm thick, confirming the few-layer structure of MXene (Figure 2(d)). The XRD pattern in Figure 2(d) reveals that $Ti_3AlC_2$ have transformed into $Ti_3C_2T_x$, while the XRD pattern of PVB indicates unimodal, which implies that it is a high molecular-weight polymer. EDS mapping of $Ti_3C_2T_x$ shows that its surface-terminated groups mainly contain -F, -O, and -OH (Figure S2(a-g)) and reveal the weight and atomic percentage of each element (Figure S2(h-i)). Because the prepared MXene nanosheets have a less-layered structure, no further treatment process is required and the aqueous MXene solution is directly used to prepare the MXene/PVB-based sensor. As can be seen from Figure 2(d), the AFM image of $Ti_3C_2T_x$ indicates that the MXene nanosheets are ~2.391 nm thick.

The MXene/PVB-based sensor consists of four layers (Figure 2(f)), the bottom layer is the interdigital electrode on the PET substrate, the middle layer contains MXene and the PVB layers, and the top layer is the insulating tape layer. Enlarged SEM images of the MXene and PVB layers are shown in Figure 2(g); more layered MXene structures are available in Figure S3. In order to study the superior mechanical properties of the MXene/PVB sensor, the structure of PVB sprayed onto MXene is compared with that of pure PVB, and Figure 2(h) reveals that this PVB has a folded structure that is more wear-resisting than the porous structure of PVB (Figure 2(i)). In order to characterize the thickness of the MXene and PVB layer after spraying (Figure S3), 5 mL and 10 mL aqueous 2.1 mg/mL MXene solutions are separately sprayed onto PET substrates containing the interdigital finger circuit. The MXene layers are found to be ~45.02 μm and ~78.75 μm thick, respectively after spraying. Spraying a solution of 1 g of PVB in 10 mL of ethanol onto the PET substrate results in a ~238.2-μm-thick PVB layer.

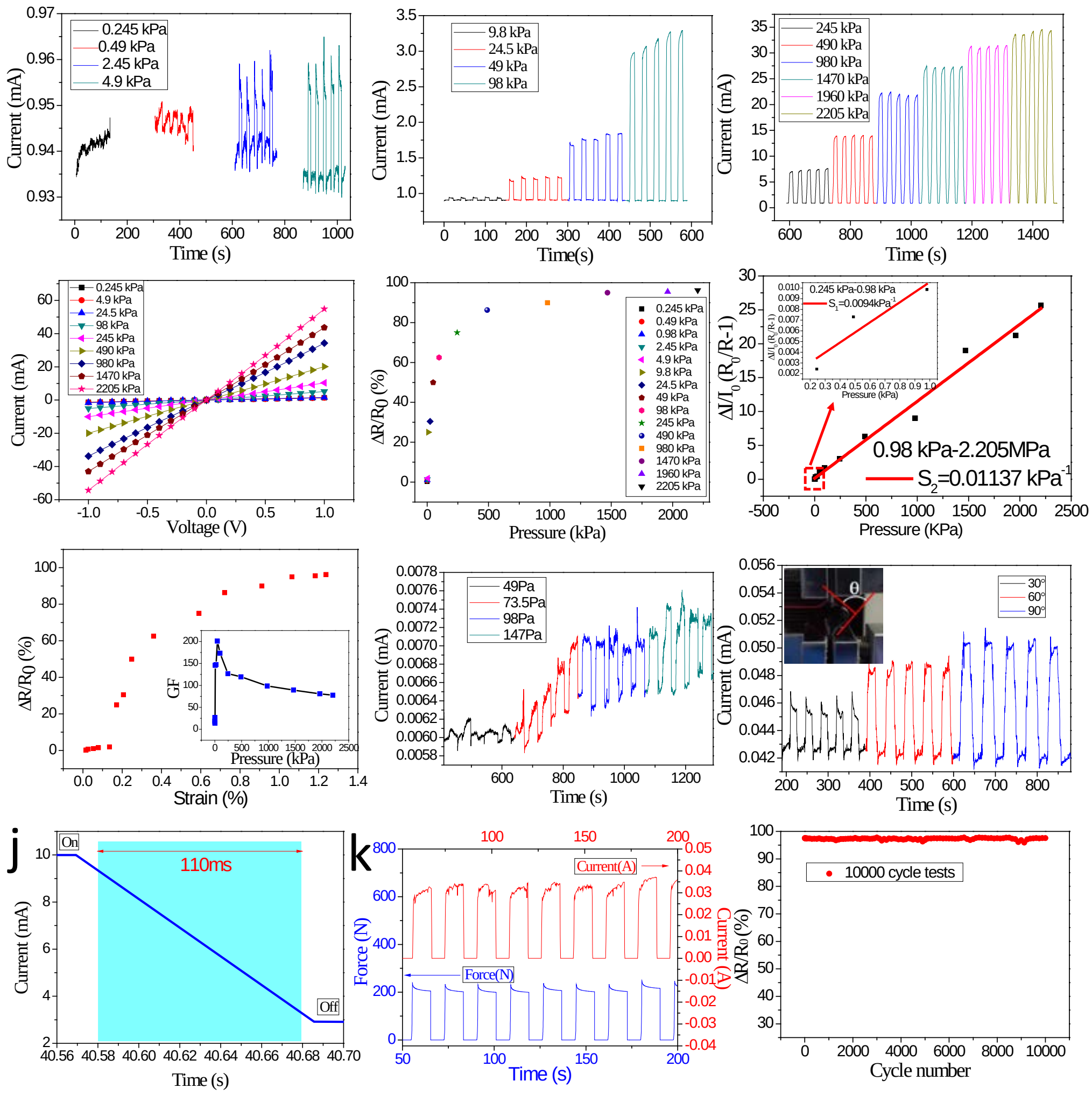

**Figure 3.** Properties of the MXene/PVB-based piezoresistive sensor. (a-c) I-T curves at different pressures. (d) I-V curves at different pressures. (e) $\Delta R/R_0$ as a function of pressure. (f) Linear fittings of the $\Delta I/I_0$ curve at different pressures. The insert shows the linear fittings at low pressure range. (g) $\Delta R/R_0$ as a function of strain. The insert shows GF as a function of pressure. (h) I-T curves at the minimum pressure detectable by the sensor. (i) I-T curves at different bending angles. (j) I-T curves as functions of response time. (k) The output current and external pressure are well synchronized with loading and unloading. (l) Mechanical durability testing of the MXene/PVB-based sensor with a maximum loading and unloading of ~2 MPa over 10,000 cycles.

The currents at different pressures is measured in order to study the piezoresistance of the MXene/PVB-based sensor. Figure S4 (a) shows the universal pressure machine and associated source meter used to test sensor performance, with the testing principle shown in Figure S4(b). Figure 3(a-c) displays I-T curves at various pressures, which demonstrate that the current increases with increasing pressure. Each pressure has been repeatedly loaded and unloaded for five times, and it is found that the current performance is clearly stable and continuous with no obvious signal attenuation during loading and unloading. The linear relationship between -1 and 1 V in the I-V curves in Figure 3(d) indicates that resistance is almost constant at each pressure. The slope of the I-V curve increases with increasing pressure, which indicates that the sensor resistance decreases. The relative change in resistance ($\Delta R/R_0$) with increasing pressure is measured, the results of which are shown in Figure 3(e). The increase in $\Delta R/R_0$ is accompanied by stress, as exhibited through three clearly different processes: a sharp rise in $\Delta R/R_0$ is observed under 49 KPa, a relatively rapid increase is observed between 49 KPa and 490 KPa, while $\Delta R/R_0$ rises relatively slowly between 490 KPa and 2.205 MPa. Due to the limits of the stretcher used in this study, the maximum testable working pressure is ~2 MPa. At a relatively small pressure, $\Delta R/R_0$ changes more quickly; however, the MXene interlayer distance becomes increasingly difficult to compress as the pressure is increased further to reach a state of almost complete conduction. These observations can be understood based on the schematic diagram in Figure 1(a-b), which shows that the parallel resistance would reach a maximum saturation value, because after that no more resistance can contribute to the parallel connectivity. For example, the resistance changes from ~200 Ω to ~7–8 Ω as the pressure reaches ~2 MPa, which indicates that the MXene layers are almost compressed to the maximum limit.

As mentioned above, the resistance of the sensor decreases with increasing pressure. Due to the large interlayer spacings between the lamellar structure layers of MXene, the amount of compression is relatively large under at 49 KPa, with a relatively large change in resistance observed, while the change in resistance is relatively small at high pressures that exceed 490 KPa. The linear fittings shown in Figure 3(f) reveal that the relationship between $\Delta R/R_0$ and pressure can be divided into two parts : below ~ 0.98 KPa, where a rate of $S_1$(0.0094 $KPa^{-1}$ , the insert figure in Figure 3(f)) is determined, and between ~490 KPa and 2.205 MPa, where the rate is ~0.01137 $KPa^{-1}$. The rate of entire pressure range is ~0.01 $KPa^{-1}$ which shows that there is a very wide linear region.

Gauge factor (GF) is a sensor coefficient that is usually used to represent the sensitivity of a sensing material, and is expressed by following formula:

$$GF = \frac{\Delta R/R_0}{\varepsilon} \tag{3}$$

where $R$ is the resistance at load, $R_0$ is the unloaded resistance, ΔR= $|R-R_0|$, and $\varepsilon$ is the strain rate. Alternatively, sensitivity can be expressed as [29]:

$$S = \frac{\delta(\Delta I/I_0)}{\delta P} \tag{4}$$

Specifically:

$$\delta\left(\frac{\Delta I}{I_0}\right) = \delta\frac{(I-I_0)}{I_0} = \delta\left(\frac{\frac{U}{R}-\frac{U}{R_0}}{\frac{U}{R_0}}\right) = \delta\left(\frac{R_0\text{-R}}{R}\right) = \delta\left(\frac{R_0}{R}-1\right) \tag{5}$$

$$\Delta I/I_0 = A \times P \qquad \text{(here } A \text{ is a constant)} \tag{6}$$

So:

$$S = \frac{\delta(\Delta I/I_0)}{\delta P} = A \quad (\text{kPa}^{-1}) \tag{7}$$

where $\Delta I$ is the relative change in current, $I_0$ is the unloaded current value, and $\Delta$P (KPa) is the pressure change between the unloaded and loaded states. As shown in Figure 3(g), GF is 20–200 below 49 KPa and 100–200 between ~49 KPa and ~2 MPa, which are significantly better values than those of other carbon-based materials, metal nanowires, and two-dimensional $MoS_2$, among others (Table S1) [24-30, 47-51]. Figure 3(g) shows that the present MXene/PVB sensor is the most sensitive in the range ~98 KPa with a GF of up to 202.2. Clearly, different linear relationships exist at different pressure intervals. Figure 3(h) shows that the minimum pressure detectable by the sensor is ~49 Pa. Clearly, the quality of the MXene is a key factor that determines sensor sensitivity. The relationship between pressure and strain during a single cycle is shown in Figure S5(a). The I-T curve in Figure S5(b) shows pressure slowly increasing from 0 to ~440 N. In this work, the prepared sensor electrode covers a PET substrate;

hence, it is very flexible and can bend and deform. The response characteristics of the MXene/PVB sensor at different bending angles were also studied, the results of which are displayed in Figure 3(i). The sensor is subject to bending deformations of 30°, 60°, and 90° under the action of testing machine; the resulting I-T curve is shown in Figure 3(i), which reveals that larger bending angles exhibit higher currents. This observation is convenient for measuring human movement and subtle changes in stress. The sensor quickly returns to its original state without degradation following bending to large deformation angles.

The response time of the sensor is determined to be ~110 ms using hand contact and release, as shown in Figure 3(j), which indicates that the sensor is able to respond quickly and recover quickly. Figure 3(k) reveals that the current and pressure are changed synchronously; hence the output current provides a good measure of loading and unloading behaviors. In order to test the mechanical durability of the MXene/PVB-based sensor, its performance over 10,000 cycles, with a maximum loading of ~2.205 MPa and a minimum loading of 0 MPa, is examined, which reveals very little attenuation of the sensor signal, as shown in Figure 3(l), with a $\Delta R/R_0$ value above 95% during 10,000 cycles. The resistance of the sensor is slightly larger after the long period of mechanical endurance testing. The minimum resistance at maximum loading also increases during cycling. Loading and unloading is performed for 5000 cycles at the maximum pressure of 2.205 MPa with the current observed to decrease during cycling (Figure S5 (c-f)), which is similar to observations made for other sensors [24,47,48, 52]. Little attenuation (<10%) is observed under such an intense pressure for 5000 cycles of loading and unloading indicative of high robustness. It is found that the MXene/PVB-based sensor is very sensitive and highly stable (see Movies S1 and S2). Figure S6 shows the I-T curve acquired at 0.1 mV, which reveals that the power consumption of the sensor is ~$3.6 \times 10^{-10}$ W, which is low. Thus, the MXene/PVB-based sensor can operate effectively at an input voltage of 0.1 mV with a current of the order of $10^{-6}$A, which corresponds to a power consumption of ~$10^{-10}$ W. In order to further test the performance of the sensor, we connect the sensor in series with a luminous LED to construct a circuit, the results of which clearly demonstrate the high sensitivity of the sensor under the action of an external force, as shown in Figure S7 (or watch movie S1 and S2). Figure S7 (a-f) shows that the circuit resistance decreases and the lamp becomes brighter with increasing pressure, while Figure S7(f-l) shows that the circuit resistance increases and the lamp becomes darker with decreasing pressure.

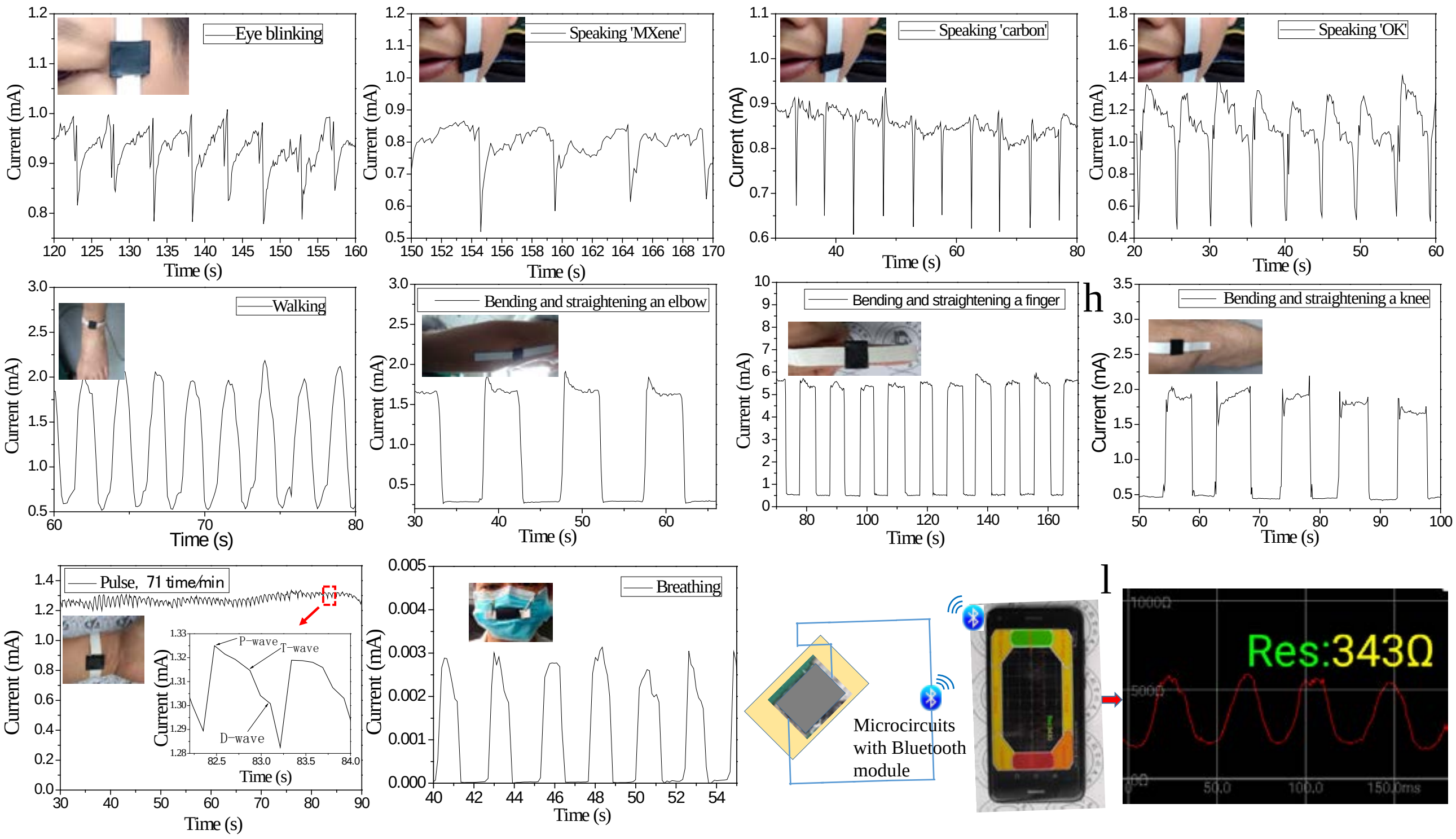

**Figure 4.** Applying the MXene/PVB-based sensor to human and other activities. I-T curves for (a) eye blinking and speaking the words (b-d) ‘MXene’, ‘Carbon’ and ‘OK’. I-T curves for human activities: (e) walking (sensor attached to ankle), (f-h) bending and straightening of an elbow, finger, and a knee. (i) I-T curve for a human pulse, showing signals for the P, T, and D waves. (j) I-T curve for breathing. (k-l) The sensor connected to a series of micro-circuits and implanted into a Bluetooth system, which changes resistance in response to wireless electromagnetic signals.

The MXene/PVB-based sensor has the advantages that include good flexibility, high sensitivity, a wide detection limit, low voltage detection, and low power consumption, among others, which engenders it with a wide range of application prospects for wearable devices that monitor and human activity. The sensor is attached to the corner of an eye and the corner the mouth to monitor subtle movements caused by changes associated with blinking and speaking. The precise changes in the current flowing through the sensor reflect the blinking and the sounds made through speaking, as shown in Figure 4(a) and Figure 4(b-d), respectively, which show that different spoken words(‘MXene’, ‘carbon’ and ‘OK’) give different waveforms; the relevant I-T curves are different and can be distinguishable by comparing the shapes and intensity changes of the traces. Moreover, we place the sensor on an ankle to monitor walking, on an elbow to monitor bending and straightening, on the knuckles to monitor finger bending and straightening, and on the knee to monitor bending and straightening, respectively, as shown in the Figure. 4(e-h), which produce regular changes in the respective current curves. In order to detect even weaker changes, we teste the ability of the sensor to detect a pulse and observe pulse signals for the P, T, and D waves, as shown in Figure 4(i). In addition, we place a sensor in a mask to detect blown air, and observe regular changes in current. As shown in Figure 4(j), the shape of the I-T curve is found to be dependent upon the blowing force and frequency. In a similar manner, we also use a pen to write words on the sensor, such as ‘MXene’, ‘Carbon’, ‘OK’, with the corresponding I-T curves shown in Figure S8(a-c). The above results show that the sensor responds quickly during loading and unloading, with the current almost unchanged during the same motion; hence, the MXene/PVB-based sensor is very flexible and highly sensitive, and can be used to accurately detect a variety of motions. For portable applications, the sensor is connected to a series of micro-circuits and implanted into a Bluetooth system that converts different currents or resistances into wireless electromagnetic signals. As shown in Figure 4(k-l), the portable micro-circuit Bluetooth MXene/PVB-based sensor also responds quickly and clearly to pressure when touched by hand. When the sensor is pressed and released, the changes in resistance with time revealed a resistance of 343 Ω at a certain time, as shown in Figure 4(l).

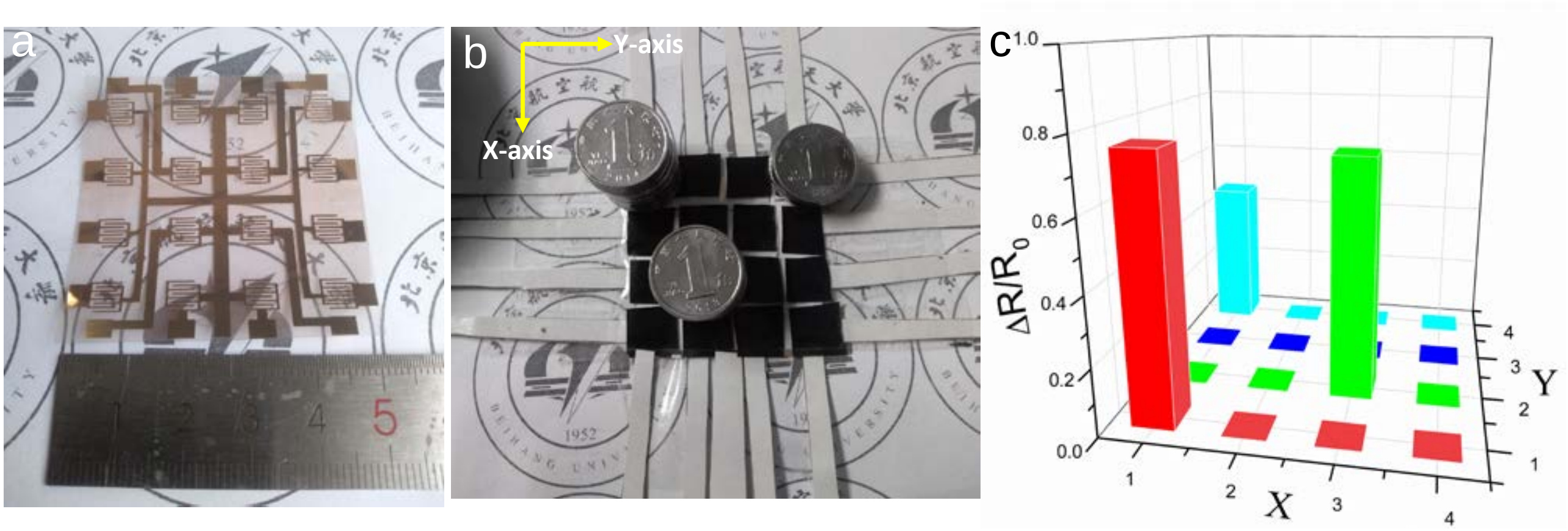

**Figure 5.** The as-fabricated 4×4 MXene/PVB-based sensor array. (a) A 4×4 sensor interdigital electrode 5 cm × 5 cm in size. (b) Stacked coins 15, 10, 5, and 0.1 RMB in value were placed at matrix positions 11, 32 and 14, respectively. (c) The corresponding pressure at each pixel was measured, with the height of each bar corresponding to the measured relative change in pressure.

As a proof of concept, we also design a 4 × 4 MXene/PVB-based sensor array, as shown in Figure 5(a-b). Each component of the sensor array is a single independent sensing unit. Stacks of coins 15, 10, 5, and 0.5 RMB in value are placed in different positions on the array, and the pressure at each location is measured. As shown in Figure 5(c), it if found that the heights of the bars that map the local pressure distribution are consistent with the positions of the coins in Figure 5(b). Hence, the MXene/PVB-based portable flexible piezoresistive sensor is potentially suitable for wearable devices and electronic skins.

## CONCLUSIONS

We have developed a simple, low-cost, ease fabrication and processing, efficient, and highly sensitive pressure sensor made up of MXene and PVB materials. The basic principle of the piezoresistive sensor involves a mechanism in which total resistance decreases with increasing parallel resistance. The MXene/PVB-based sensor exhibits a sensitivity of GF ~202.2, a response time of 110 ms, high stability (over more than 10,000 cycles of loading and unloading), and a power consumption of ~$3.6 \times 10^{-10}$ W at an input voltage of 0.1 mV. The present sensor can detect dynamic forces over a wide pressure range (49 Pa to ~2.205 MPa), and can distinguish a variety of complex forces, including pressure, bending, and acoustic vibrations. These properties facilitate the monitoring of the strengths of real signals from radial pulses and acoustic vibrations in real time. It is important to highlight that the entire fabrication process for the sensor is scalable and does not require complex and/or expensive equipment. Designing such a high-performance piezoresistive sensor with a wide pressure detection range and low power consumption, combined with suitable flexible sensor components, simplifies the fabrication steps, which opens up a new ease-of-fabrication and low-power flexible sensing avenue. To our knowledge, we believe that our present approach enables the fabrication of low-cost next-generation pressure sensors with a wide detection range that can easily be integrated into future wearable electronics, such as flexible touch screens, human-machine interface instruments, and prosthetic skin.

## EXPERIMENTAL METHODS

***Chemicals***. The $Ti_3AlC_2$ MAX-phase powder (200 mesh) and LiF (Alfa, 98+%) were purchased from the Beijing Wisdom Company and the Aladdin Reagent Company, respectively. Hydrochloric acid (reagent, 36-38%) was purchased from Beijing Chemical Works (China). Deionised (DI) water was used to prepare all solutions.

***Synthesis of MXene***. LiF (1 g) is slowly added to hydrochloric acid (15 mL) in add 5mI of deionised water, after which , $Ti_3AlC_2$ (1 g) was slowly added. The mixture is subsequently reacted for 24 h with stirring (3500 rpm) at 35 °C. A MXene solution is finally obtained by centrifugation spinning at 3500 rpm for 5 min each cycle until the pH is greater than 6. In order to obtain highly dispersed MXene nanosheets, the supernatant solution is poured from the centrifuge tube into a flask and the precipitate is manually shaken for 5 minutes. The dispersed suspension is then centrifuged at 3500 rpm for 10 min, and the resulting supernatant is collected for further testing and characterisation. The concentration of MXene is determined by weighing a certain amount of the MXene suspension, and then weighing the solid after vacuum drying.

***Fabricating the interdigital electrodes and the sensor array electrode.*** Copper clad polymer films on a 100-μm-thick PET substrate with a 200-nm-thick copper coating layer was purchased from a commercial source (Teonex Q65FA, Graphene Testing and Sales Platform Co. Ltd., Taiwan). The copper layer is coated on the flexible polymer substrate by evaporation or sputtering deposition. The interdigital electrode and sensor array electrode are fabricated by laser irradiating the film with a 3W diode-pumped Nd: YAG laser.

***Characterization***. The morphology and thickness of the MXene/PVB-based sensor were characterized by FE-SEM (ZEISS SUPRA55, Germany). Optical microscopy images of the interdigital electrode are obtained using a transparent reflectance polarising microscope (59XC-PC, Shanghai optical Instrument Factory, Shanghai, China). XRD patterns of MAX, MXene, and PVB are recorded using an X-ray diffractometer (D8 ADVANCE, Brooker, Germany). To test the response of our pressure sensor to static and dynamic pressures, we use a computer-controlled stepper motor system, a force sensor (ZQ 990B, Dongguan Zhitou Precision Instrument Co., Ltd., Guangdong, China), and a Keithley 2400 source meter (Teck Technology Co., Ltd., USA). The input voltage is set to 100 mV and 0.1 mV during testing.

## ASSOCIATED CONTENT

### Supporting Information

The Supporting Information is available free of charge on the ACS publication website at DOI:

Further experimental characterization (SEM, EDS Mapping images) of MXene nanosheets; optical images of PET-based copper interdigital electrode; SEM images of the thickness of MXene layer and PVB layer; schematic diagram of the pressure sensor experimental setup and the testing principle; mechanical properties of MXene/PVB-based sensor; sensing performances of MXene /PVB -based sensor; comparison of sensor performance based on MXene and other material. (PDF)

## AUTHOR INFORMATION

Corresponding authors:
*E-mail:gcshan@buaa.edu.cn.
*iamwhuang@nwpu.edu.cn.

## CONFLICT OF INTEREST

The authors declare no conflicts of interest.

## ACKNOWLEDGEMENT

The work was supported by the National Key R&D Program of China (Grant No. 2016YFE0204200), and the National Natural Science Foundation of China (NSFC Grant No. 51702009).

## REFERENCES

[1] Naguib, M.; Kurtoglu, M.; Presser, V.; Lu, J.; Niu, J.; Heon, M.; Hultman, L.; Gogotsi, Y.; Barsoum, M. W. Two-Dimensional Nanocrystals Produced by Exfoliation of Ti 3AlC 2. *Adv. Mater.* **2011**, *23* (37), 4248–4253.

[2] Naguib, M.; Mochalin, V. N.; Barsoum, M. W.; Gogotsi, Y. 25th Anniversary Article: MXenes: A New Family of Two-Dimensional Materials. *Adv. Mater.* **2014**, *26* (7), 992–1005.

[3] Jiang, Q.; Wu, C.; Wang, Z.; Wang, A. C.; He, J. H.; Wang, Z. L.; Alshareef, H. N. MXene Electrochemical

Microsupercapacitor Integrated with Triboelectric Nanogenerator as a Wearable Self-Charging Power Unit. *Nano Energy* **2018**, *45* (January), 266–272.

[4] Chang, T. H.; Zhang, T.; Yang, H.; Li, K.; Tian, Y.; Lee, J. Y.; Chen, P. Y. Controlled Crumpling of Two-Dimensional Titanium Carbide (MXene) for Highly Stretchable, Bendable, Efficient Supercapacitors. *ACS Nano* **2018**, *12* (8), 8048–8059.

[5] Wang, Z.; Qin, S.; Seyedin, S.; Zhang, J.; Wang, J.; Levitt, A.; Li, N.; Haines, C.; Ovalle-Robles, R.; Lei, W.; et al. High-Performance Biscrolled MXene/Carbon Nanotube Yarn Supercapacitors. *Small* **2018**, *14* (37), 1–9.

[6] Zhang, Q.; Wang, F.; Zhang, H.; Zhang, Y.; Liu, M.; Liu, Y. Universal Ti 3 C 2 MXenes Based Self-Standard Ratiometric Fluorescence Resonance Energy Transfer Platform for Highly Sensitive Detection of Exosomes. *Anal. Chem.* **2018**, *90* (21), 12737–12744.

[7] Chertopalov, S.; Mochalin, V. N. Environment-Sensitive Photoresponse of Spontaneously Partially Oxidized Ti3C2 MXene Thin Films. *ACS Nano* **2018**, *12* (6), 6109–6116.

[8] Deng, W.; Huang, H.; Jin, H.; Li, W.; Chu, X.; Xiong, D.; Yan, W.; Chun, F.; Xie, M.; Luo, C.; et al. All-Sprayed-Processable, Large-Area, and Flexible Perovskite/MXene-Based Photodetector Arrays for Photocommunication. *Adv. Opt. Mater.* **2019**, *7* (6), 1–9.

[9] Xu, B.; Zhu, M.; Zhang, W.; Zhen, X.; Pei, Z.; Xue, Q.; Zhi, C.; Shi, P. Ultrathin MXene-Micropattern-Based Field-Effect Transistor for Probing Neural Activity. *Adv. Mater.* **2016**, *28* (17), 3333–3339.

[10] Lee, E.; Vahidmohammadi, A.; Prorok, B. C.; Yoon, Y. S.; Beidaghi, M.; Kim, D. J. Room Temperature Gas Sensing of Two-Dimensional Titanium Carbide (MXene). *ACS Appl. Mater. Interfaces* **2017**, *9* (42), 37184–37190.

[11] Kim, S. J.; Koh, H. J.; Ren, C. E.; Kwon, O.; Maleski, K.; Cho, S. Y.; Anasori, B.; Kim, C. K.; Choi, Y. K.; Kim, J.; et al. Metallic Ti3C2Tx MXene Gas Sensors with Ultrahigh Signal-to-Noise Ratio. *ACS Nano* **2018**, *12* (2), 986–993.

[12] Chen, X.; Sun, X.; Xu, W.; Pan, G.; Zhou, D.; Zhu, J.; Wang, H.; Bai, X.; Dong, B.; Song, H. Ratiometric Photoluminescence Sensing Based on Ti3C2 MXene Quantum Dots as an Intracellular PH Sensor. *Nanoscale* **2018**, *10* (3), 1111–1118.

[13] Chen, X.; Li, J.; Pan, G.; Xu, W.; Zhu, J.; Zhou, D.; Li, D.; Chen, C.; Lu, G.; Song, H. Ti 3 C 2 MXene Quantum Dots/TiO 2 Inverse Opal Heterojunction Electrode Platform for Superior Photoelectrochemical Biosensing. *Sensors Actuators, B Chem.* **2019**, *289* (November 2018), 131–137.

[14] Kumar, S.; Lei, Y.; Alshareef, N. H.; Quevedo-Lopez, M. A.; Salama, K. N. Biofunctionalized Two-Dimensional Ti3C2 MXenes for Ultrasensitive Detection of Cancer Biomarker. *Biosens. Bioelectron.* **2018**, *121* (August), 243–249.

[15] Liu, J.; Jiang, X.; Zhang, R.; Zhang, Y.; Wu, L.; Lu, W.; Li, J.; Li, Y.; Zhang, H. MXene-Enabled Electrochemical Microfluidic Biosensor: Applications toward Multicomponent Continuous Monitoring in Whole Blood. *Adv. Funct. Mater.* **2019**, *29* (6), 1–9.

[16] Muckley, E. S.; Naguib, M.; Ivanov, I. N. Multi-Modal, Ultrasensitive, Wide-Range Humidity Sensing with $Ti_3C_2$ Film. *Nanoscale* **2018**, *10* (46), 21689–21695.

[17] Muckley, E. S.; Naguib, M.; Wang, H. W.; Vlcek, L.; Osti, N. C.; Sacci, R. L.; Sang, X.; Unocic, R. R.; Xie, Y.; Tyagi, M.; et al. Multimodality of Structural, Electrical, and Gravimetric Responses of Intercalated MXenes to Water. *ACS Nano* **2017**, *11* (11), 11118–11126.

[18] Fang, Y.; Yang, X.; Chen, T.; Xu, G.; Liu, M.; Liu, J.; Xu, Y. Two-Dimensional Titanium Carbide (MXene)-Based Solid-State Electrochemiluminescent Sensor for Label-Free Single-Nucleotide Mismatch Discrimination in Human Urine. *Sensors Actuators, B Chem.* **2018**, *263*, 400–407.

[19] Shankar, S. S.; Shereema, R. M.; Rakhi, R. B. Electrochemical Determination of Adrenaline Using MXene/Graphite Composite Paste Electrodes. *ACS Appl. Mater. Interfaces* **2018**, *10* (50), 43343–43351.

[20] Yang, Y.; Shi, L.; Cao, Z.; Wang, R.; Sun, J. Strain Sensors with a High Sensitivity and a Wide Sensing Range Based on a Ti 3 C 2 T x (MXene) Nanoparticle–Nanosheet Hybrid Network. *Adv. Funct. Mater.* **2019**, *29* (14), 1–10.

[21] Sougrat, R.; Zhang, Y.-Z.; Kim, H.; Anjum, D. H.; Lee, K. H.; Alshareef, H. N.; Jiang, Q. MXenes Stretch Hydrogel Sensor Performance to New Limits. *Sci. Adv.* **2018**, No. June, 1–8.

[22] Habib, T.; Green, M. J.; Shah, S.; Radovic, M.; An, H.; Lutkenhaus, J. L.; Gao, H. Surface-Agnostic Highly Stretchable and Bendable Conductive MXene Multilayers. *Sci. Adv.* **2018**, *4* (3), eaaq0118.

[23] Shi, X.; Wang, H.; Xie, X.; Xue, Q.; Zhang, J.; Kang, S.; Wang, C.; Liang, J.; Chen, Y. Bioinspired Ultrasensitive and Stretchable MXene-Based Strain Sensor via Nacre-Mimetic Microscale “Brick-and-Mortar” Architecture. *ACS Nano* **2019**, *13* (1), 649–659.

[24] Ma, Y.; Liu, N.; Li, L.; Hu, X.; Zou, Z.; Wang, J.; Luo, S.; Gao, Y. A Highly Flexible and Sensitive Piezoresistive Sensor Based on MXene with Greatly Changed Interlayer Distances. *Nat. Commun.* **2017**, *8* (1), 1–7.

[25] Yue, Y.; Liu, N.; Liu, W.; Li, M.; Ma, Y.; Luo, C.; Wang, S.; Rao, J.; Hu, X.; Su, J.; et al. 3D Hybrid Porous Mxene-Sponge Network and Its Application in Piezoresistive Sensor. *Nano Energy* **2018**, *50* (May), 79–87.

[26] Guo, Y.; Zhong, M.; Fang, Z.; Wan, P.; Yu, G. A Wearable Transient Pressure Sensor Made with MXene Nanosheets for Sensitive Broad-Range Human-Machine Interfacing. *Nano Lett.* **2019**, *19* (2), 1143–1150.

[27] Li, X. P.; Li, Y.; Li, X.; Song, D.; Min, P.; Hu, C.; Zhang, H. Bin; Koratkar, N.; Yu, Z. Z. Highly Sensitive, Reliable and Flexible Piezoresistive Pressure Sensors Featuring Polyurethane Sponge Coated with MXene Sheets. *J. Colloid Interface Sci.* **2019**, *542*, 54–62.

[28] Li, T.; Chen, L.; Yang, X.; Chen, X.; Zhang, Z.; Zhao, T.; Li, X.; Zhang, J. A Flexible Pressure Sensor Based on an MXene-Textile Network Structure. *J. Mater. Chem. C* **2019**, *7* (4), 1022–1027.

[29] Zhuo, H.; Hu, Y.; Chen, Z.; Peng, X.; Liu, L.; Luo, Q.; Yi, J.; Liu, C.; Zhong, L. A Carbon Aerogel with Super Mechanical and Sensing Performances for Wearable Piezoresistive Sensors. *J. Mater. Chem. A* **2019**, *7* (14),

[30] Ma, Y.; Yue, Y.; Zhang, H.; Cheng, F.; Zhao, W.; Rao, J.; Luo, S.; Wang, J.; Jiang, X.; Liu, Z.; et al. 3D Synergistical MXene/Reduced Graphene Oxide Aerogel for a Piezoresistive Sensor. *ACS Nano* **2018**, *12* (4), 3209–3216.

[31] Desai, M. L.; Basu, H.; Singhal, R. K.; Saha, S.; Kailasa, S. K. Ultra-Small Two Dimensional MXene Nanosheets for Selective and Sensitive Fluorescence Detection of Ag + and Mn 2+ Ions. *Colloids Surfaces A Physicochem. Eng. Asp.* **2019**, *565* (October 2018), 70–77.

[32] Zhang, Y.; Jiang, X.; Zhang, J.; Zhang, H.; Li, Y. Simultaneous Voltammetric Determination of Acetaminophen and Isoniazid Using MXene Modified Screen-Printed Electrode. *Biosens. Bioelectron.* **2019**, *130* (October 2018), 315–321.

[33] Rasheed, P. A.; Pandey, R. P.; Rasool, K.; Mahmoud, K. A. Ultra-Sensitive Electrocatalytic Detection of Bromate in Drinking Water Based on Nafion/Ti3C2Tx (MXene) Modified Glassy Carbon Electrode. *Sensors Actuators, B Chem.* **2018**, *265*, 652–659.

[34] Zheng, J.; Wang, B.; Ding, A.; Weng, B.; Chen, J. Synthesis of MXene/DNA/Pd/Pt Nanocomposite for Sensitive Detection of Dopamine. *J. Electroanal. Chem.* **2018**, *816* (March), 189–194.

[35] Dong, Y.; Mallineni, S. S. K.; Maleski, K.; Behlow, H.; Mochalin, V. N.; Rao, A. M.; Gogotsi, Y.; Podila, R. Metallic MXenes: A New Family of Materials for Flexible Triboelectric Nanogenerators. *Nano Energy* **2018**, *44* (October 2017), 103–110.

[36] Yang, T.; Xie, D.; Li, Z.; Zhu, H. Recent Advances in Wearable Tactile Sensors: Materials, Sensing Mechanisms, and Device Performance. *Mater. Sci. Eng. R Reports* **2017**, *115*, 1–37.

[37] Chen, Z.; Wang, Z.; Li, X.; Lin, Y.; Luo, N.; Long, M.; Zhao, N.; Xu, J. Bin. Flexible Piezoelectric-Induced Pressure Sensors for Static Measurements Based on Nanowires/Graphene Heterostructures. *ACS Nano* **2017**, *11* (5), 4507–4513.

[38] Van Lier, G.; Van Alsenoy, C.; Van Doren, V.; Geerlings, P. Ab Initio Study of the Elastic Properties of Single-Walled Carbon Nanotubes and Graphene. *Chem. Phys. Lett.* **2000**, *326* (1–2), 181–185.

[39] Han, S. T.; Peng, H.; Sun, Q.; Venkatesh, S.; Chung, K. S.; Lau, S. C.; Zhou, Y.; Roy, V. A. L. An Overview of the Development of Flexible Sensors. *Adv. Mater.* **2017**, *29* (33), 1–22.

[40] Liu, S. Y.; Lian, L.; Pan, J.; Lu, J. G.; Shieh, H. P. D. Highly Sensitive and Optically Transparent Resistive Pressure Sensors Based on a Graphene/Polyaniline-Embedded PVB Film. *IEEE Trans. Electron Devices* **2018**, *65* (5), 1939–1945.

[41] Jun, S.; Han, C. J.; Kim, Y.; Ju, B. K.; Kim, J. W. A Pressure-Induced Bending Sensitive Capacitor Based on an Elastomer-Free, Extremely Thin Transparent Conductor. *J. Mater. Chem. A* **2017**, *5* (7), 3221–3229.

[42] Yuan, R.; Liu, Y.; Li, Q. F.; Chai, Y. Q.; Mo, C. L.; Zhong, X.; Tang, D. P.; Dai, J. Y. Electrochemical Characteristics of a Platinum Electrode Modified with a Matrix of Polyvinyl Butyral and Colloidal Ag Containing Immobilized Horseradish Peroxidase. *Anal. Bioanal. Chem.* **2005**, *381* (3), 762–768.

[43] Arshak, K.; Morris, D.; Arshak, A.; Korostynska, O. Sensitivity of Polyvinyl Butyral/Carbon-Black Sensors to Pressure. *Thin Solid Films* **2008**, *516* (10), 3298–3304.

[44] Korostynska, O.; Arshak, A.; Arshak, K.; Morris, D. Investigation into Real-Time Pressure Sensing Properties of SnO 2, TiO2, and TiO2/ZnO Thick Films with Interdigitated Electrodes. *Mater. Sci. Eng. B Solid-State Mater. Adv. Technol.* **2011**, *176* (16), 1297–1300.

[45] Zhu, G.; Cui, X.; Zhang, Y.; Chen, S.; Dong, M.; Liu, H.; Shao, Q.; Ding, T.; Wu, S.; Guo, Z. Poly (Vinyl Butyral)/Graphene Oxide/Poly (Methylhydrosiloxane) Nanocomposite Coating for Improved Aluminum Alloy Anticorrosion. *Polymer (Guildf).* **2019**, *172* (March), 415–422.

[46] Qin, R.; Hu, M.; Zhang, N.; Guo, Z.; Yan, Z.; Li, J.; Liu, J.; Shan, G.; Yang, J. Flexible Fabrication of Flexible Electronics: A General Laser Ablation Strategy for Robust Large-Area Copper-Based Electronics. *Adv. Electron. Mater.* **2019**, *1900365*, 1900365.

[47] Gong, S.; Schwalb, W.; Wang, Y.; Chen, Y.; Tang, Y.; Si, J.; Shirinzadeh, B.; Cheng, W. A Wearable and Highly Sensitive Pressure Sensor with Ultrathin Gold Nanowires. *Nat. Commun.* **2014**, *5*, 1–8.

[48] Wu, X.; Han, Y.; Zhang, X.; Zhou, Z.; Lu, C. Large-Area Compliant, Low-Cost, and Versatile Pressure-Sensing Platform Based on Microcrack-Designed Carbon Black@Polyurethane Sponge for Human–Machine Interfacing. *Adv. Funct. Mater.* **2016**, *26* (34), 6246–6256.

[49] Yan, C.; Wang, J.; Kang, W.; Cui, M.; Wang, X.; Foo, C. Y.; Chee, K. J.; Lee, P. S. Highly Stretchable Piezoresistive Graphene-Nanocellulose Nanopaper for Strain Sensors. *Adv. Mater.* **2014**, *26* (13), 2022–2027.

[50] Li, Y.; Luo, S.; Yang, M. C.; Liang, R.; Zeng, C. Poisson Ratio and Piezoresistive Sensing: A New Route to High-Performance 3D Flexible and Stretchable Sensors of Multimodal Sensing Capability. *Adv. Funct. Mater.* **2016**, *26* (17), 2900–2908.

[51] Park, M.; Park, Y. J.; Chen, X.; Park, Y. K.; Kim, M. S.; Ahn, J. H. MoS2-Based Tactile Sensor for Electronic Skin Applications. *Adv. Mater.* **2016**, *28* (13), 2556–2562.

[52] Bae, G. Y.; Pak, S. W.; Kim, D.; Lee, G.; Kim, D. H.; Chung, Y.; Cho, K. Linearly and Highly Pressure-Sensitive Electronic Skin Based on a Bioinspired Hierarchical Structural Array. *Adv. Mater.* **2016**, *28* (26), 5300–5306.

# Supporting Information

## A highly sensitive piezoresistive sensor based on MXene and polyvinyl butyral with wide detection limit and low power consumption

Ruzhan Qin[a], Mingjun Hu[b], Xin Li[a], Li Yan[b], Chuanguang Wu[b], Jinzhang Liu[b], Haibin Gao[c], Guangcun Shan*[a], and Wei Huang*[d]

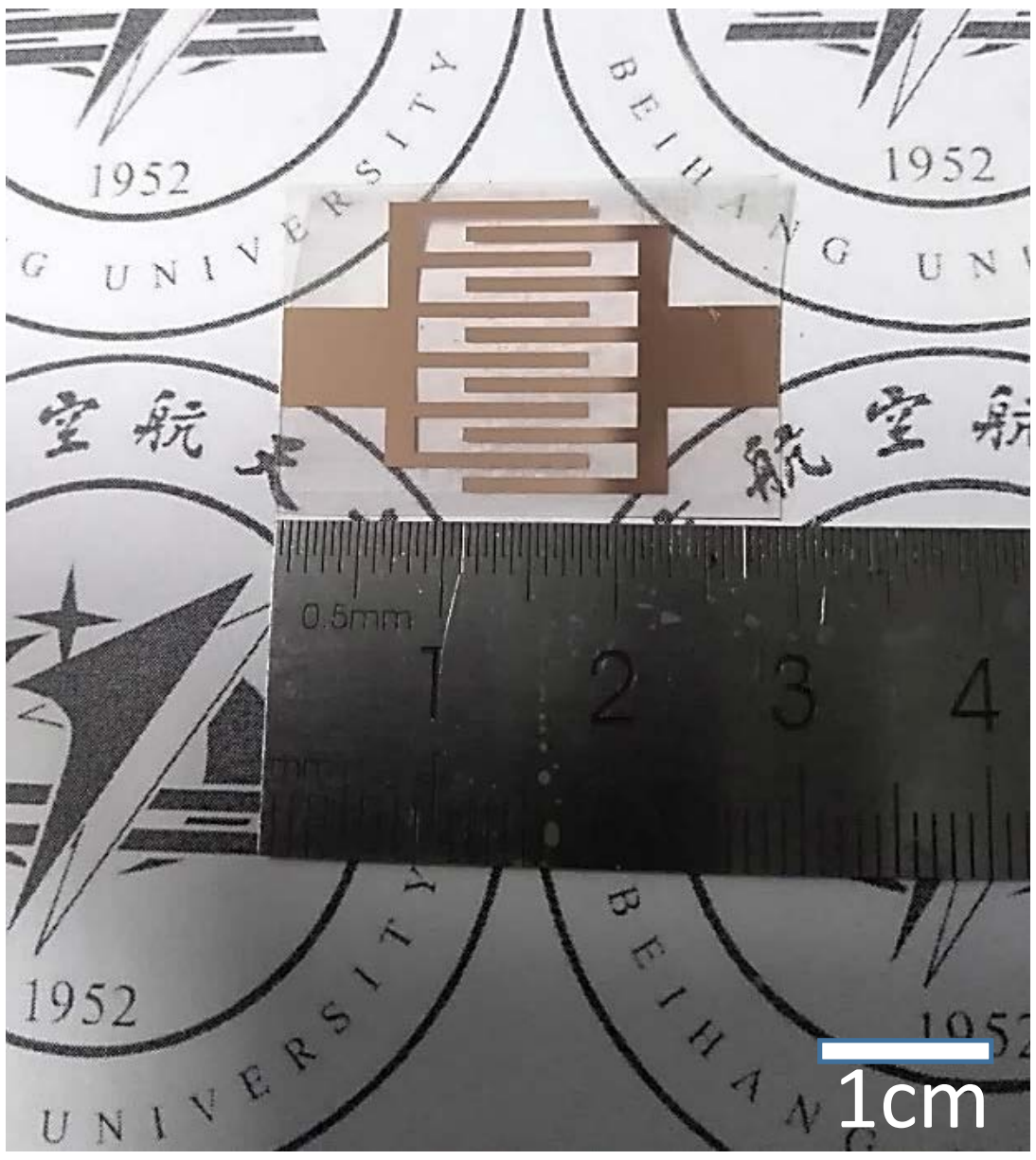

**Figure S1.** The interdigital electrode on the flexible PET substrate was prepared using our previous preparation process of laser ablation strategy.

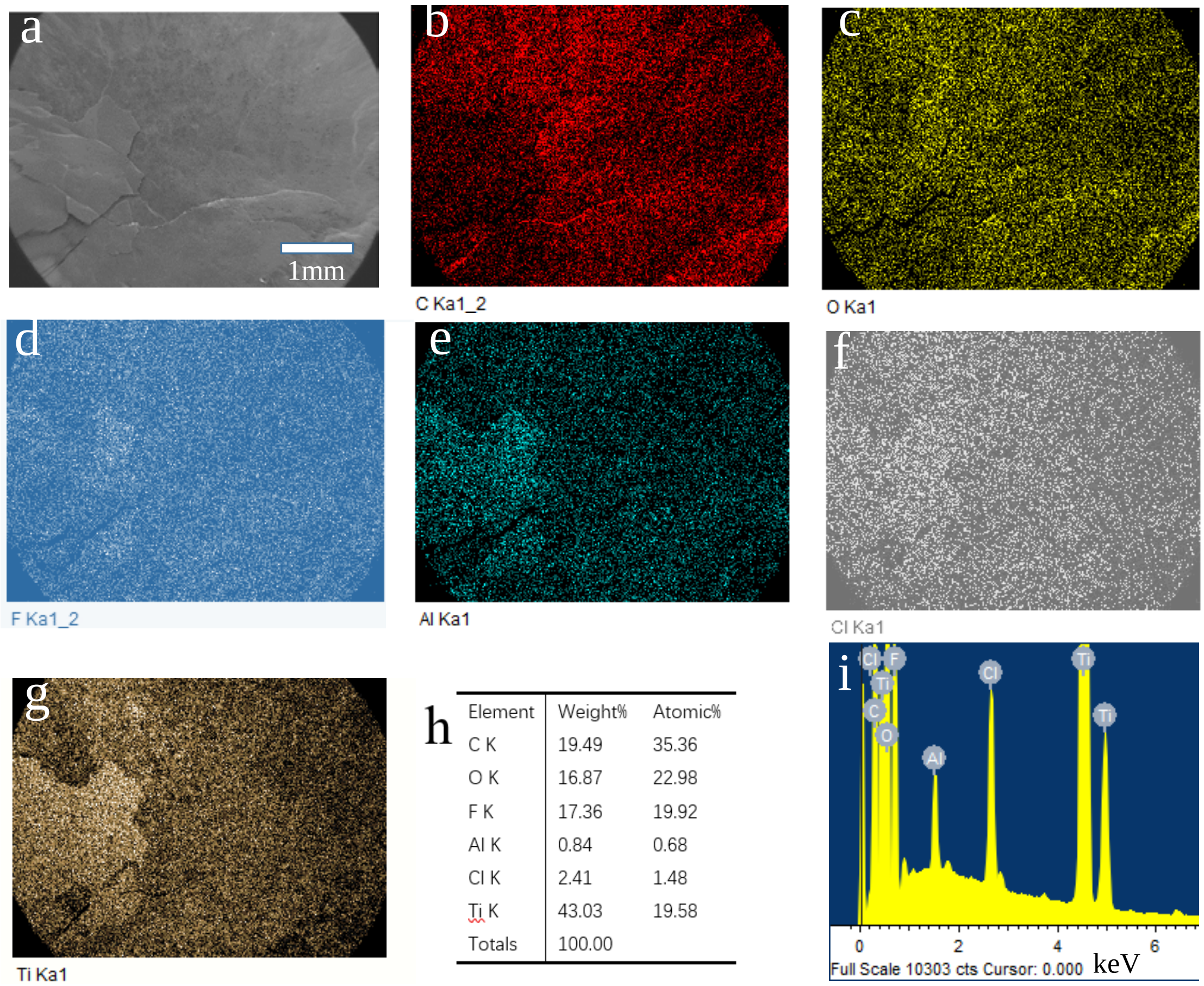

| Element | Weight% | Atomic% |
|---|---|---|
| C K | 19.49 | 35.36 |
| O K | 16.87 | 22.98 |
| F K | 17.36 | 19.92 |
| Al K | 0.84 | 0.68 |
| Cl K | 2.41 | 1.48 |
| Ti K | 43.03 | 19.58 |
| Totals | 100.00 | |

**Figure S2.** (a) SEM image of MXene layer of sensor. (b-g) EDS mapping of MXene layer of sensor. (h) The weight and atomic percentage of each element. (i) The x-ray energy dispersion spectrum(EDS).

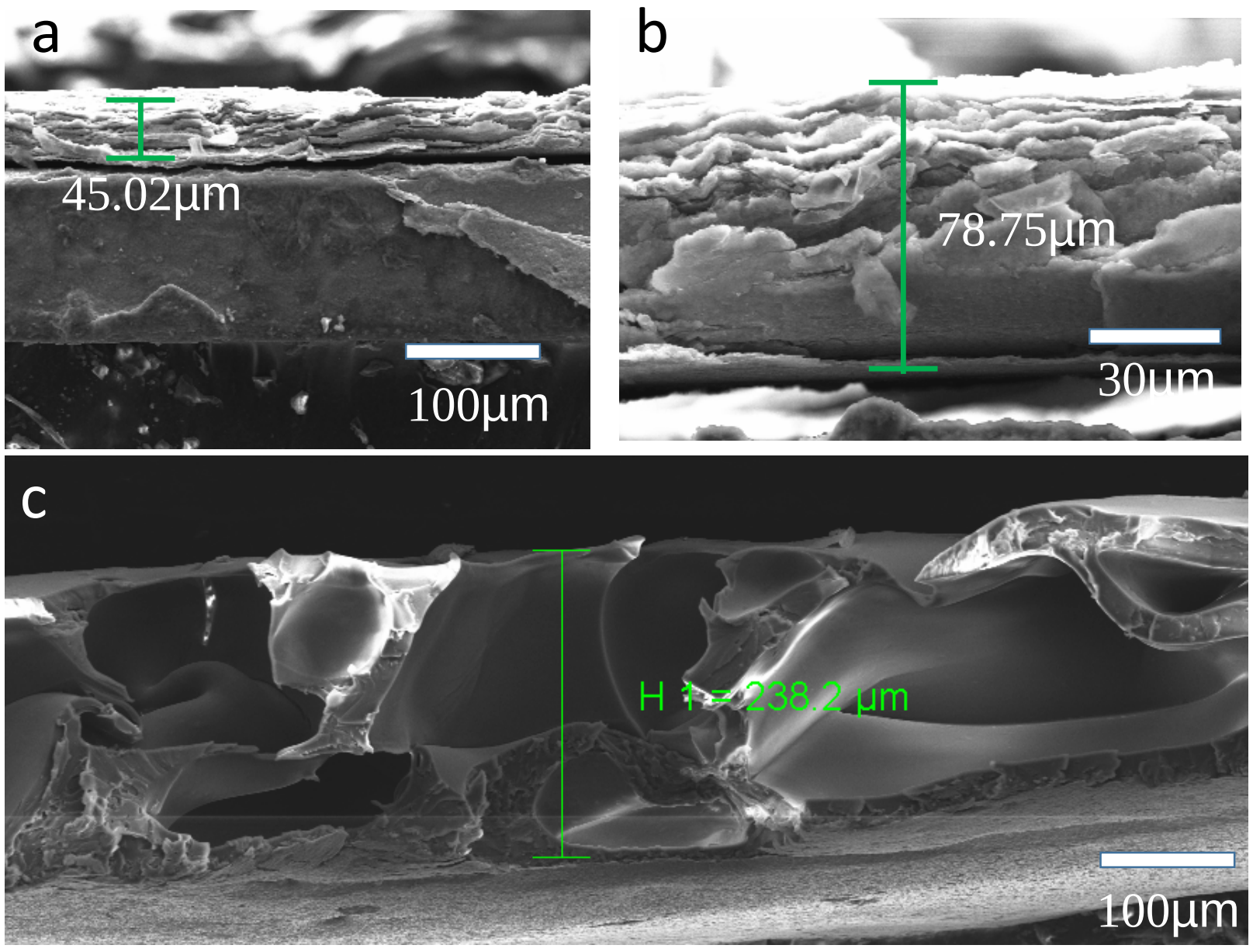

**Figure S3.** (a-b) The thickness of MXene layer after spraying 5 mL and 10 mL aqueous 2.1 mg/mL MXene solutions were separately sprayed onto PET substrates containing the interdigital finger circuit. (c) The thickness of PVB layer after Spraying a solution of 1 g of PVB in 10 mL of ethanol onto the PET substrate.

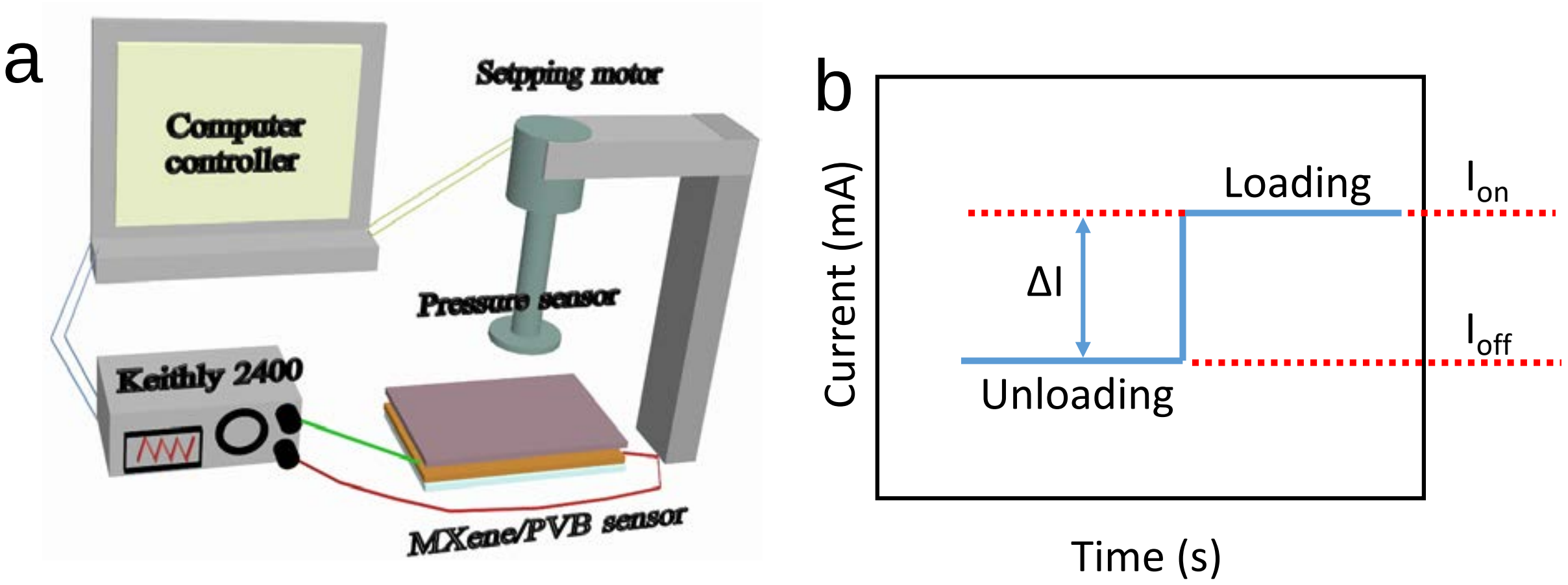

**Figure S4.** (a)The universal pressure machine and associated source meter used to test sensor performance. (b) The testing principle of sensor performance.

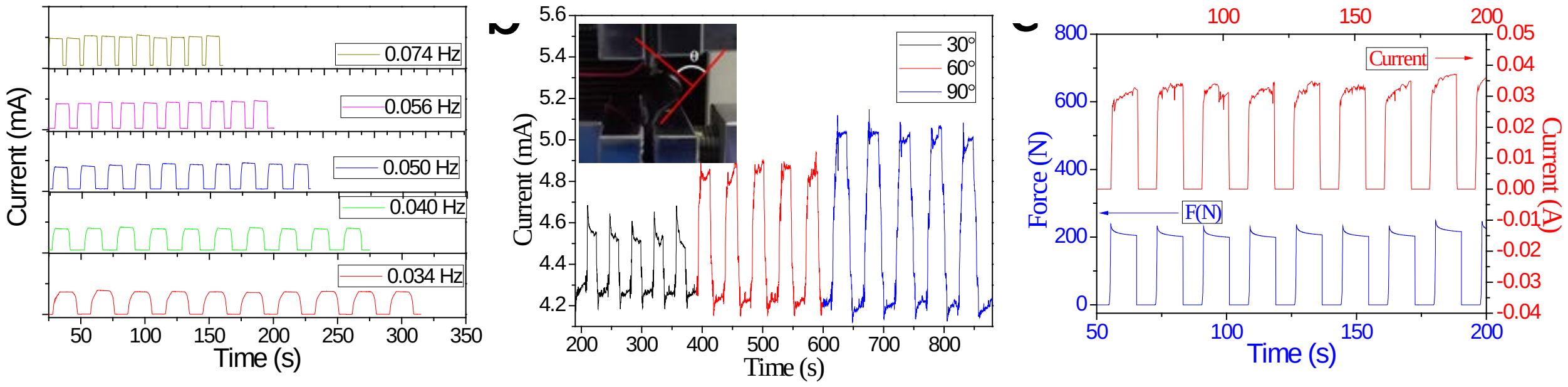

**Figure S5.** (a) I-T curves with the pressure of 1kPa at different frequency. (b) I-T curves at different bending angles. (c) The output current and external pressure are well synchronized with loading and unloading.

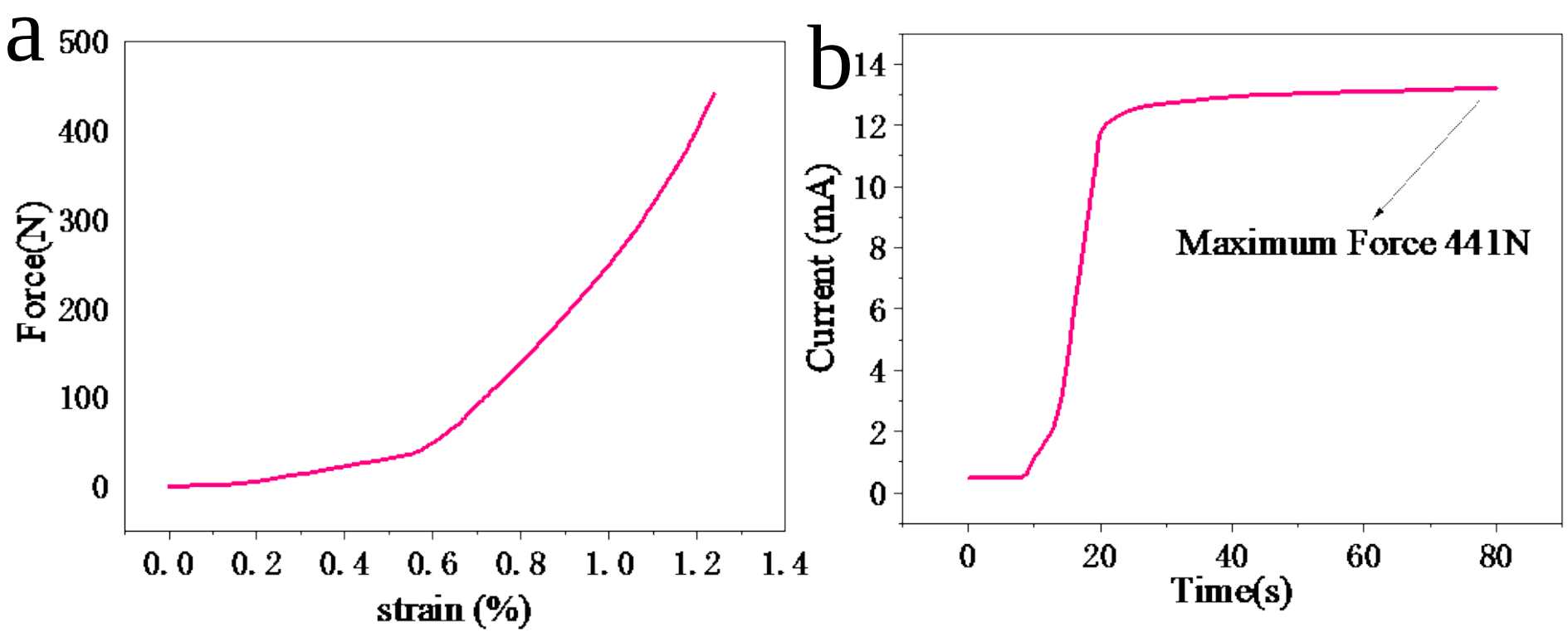

**Figure S6** (a) The relationship between pressure and strain during a single cycle. (b) The relationship between pressure and time during a single cycle.

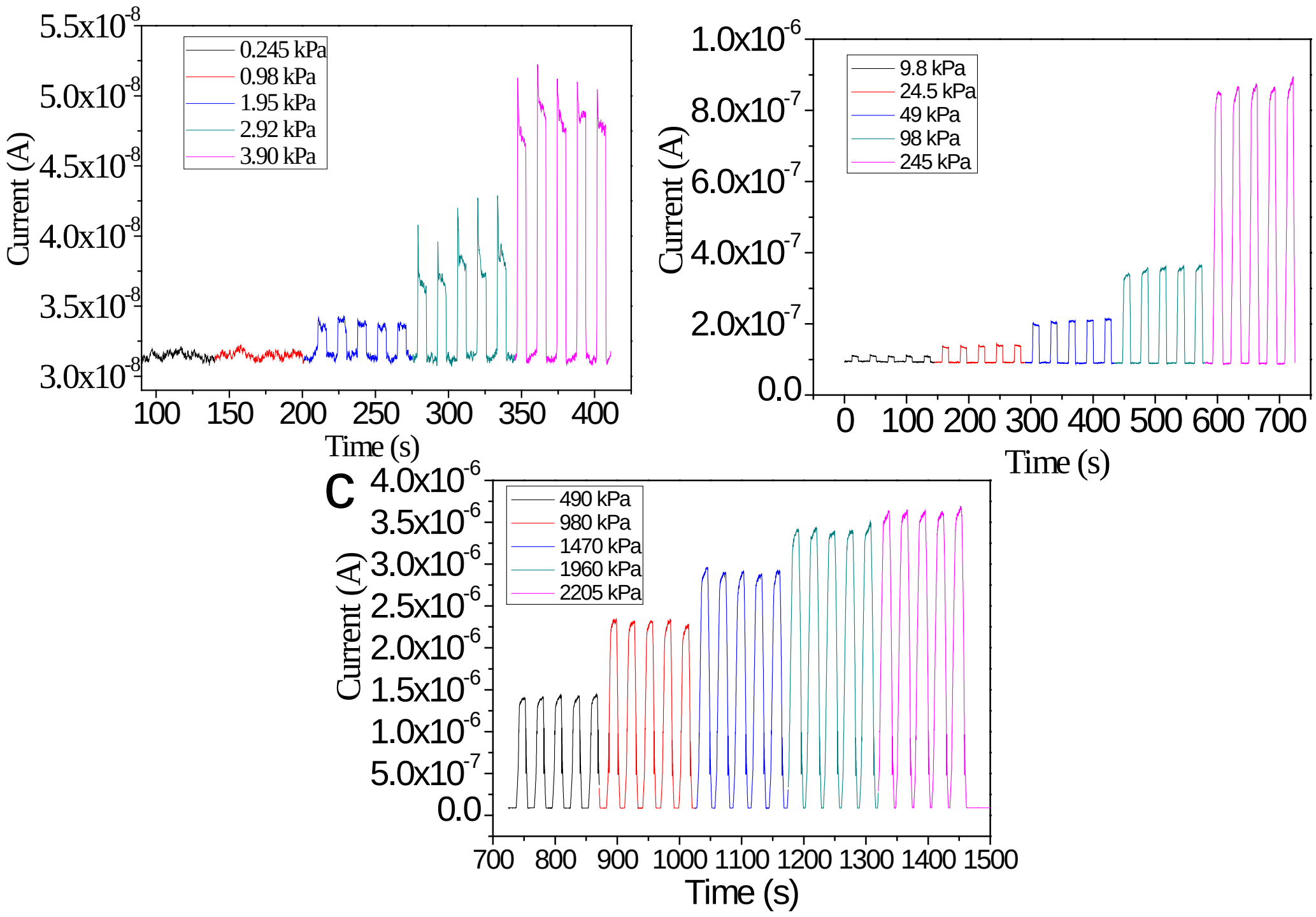

**Figure S7.** (a-c) the I-T curve acquired at 0.1 mV, which reveals that the power consumption of the sensor is ~3.6 × $10^{-10}$ W.

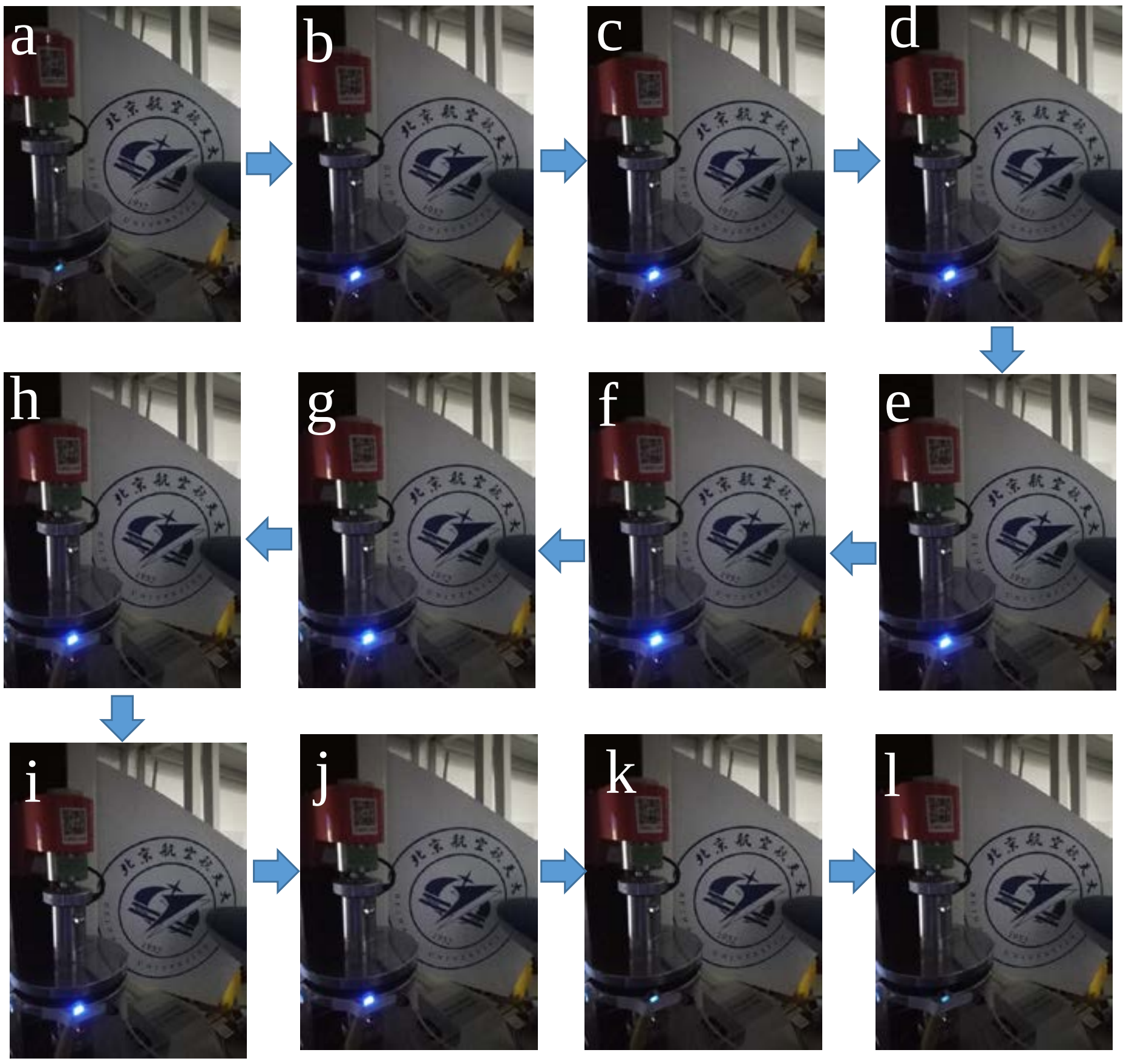

**Figure S8.** (a-f) The circuit resistance decreases and the lamp becomes brighter with increasing pressure. (g-l) The circuit resistance increases and the lamp becomes darker with decreasing pressure.

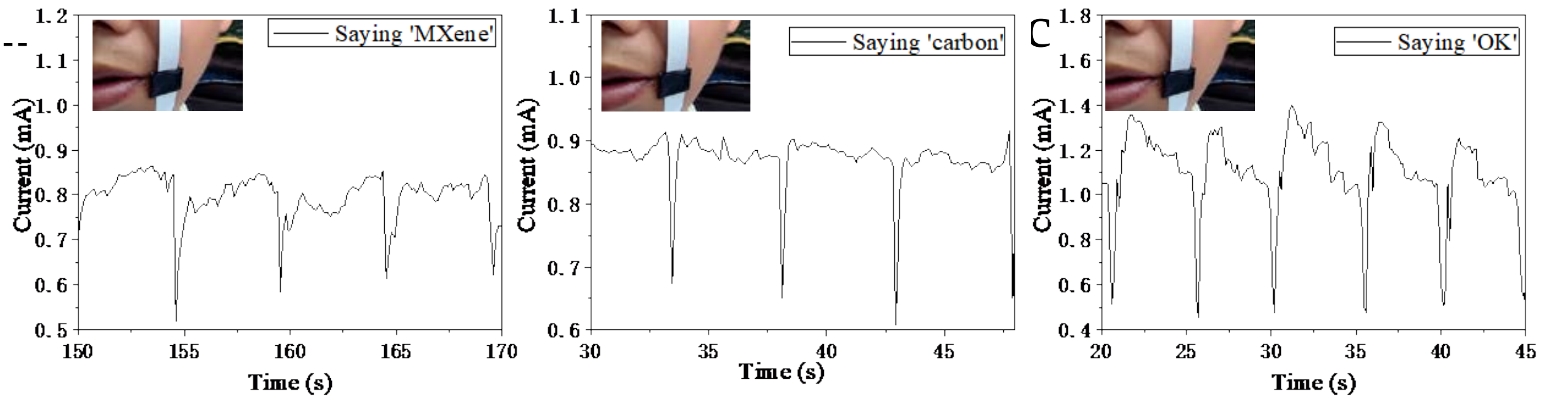

**Figure S9.** (a-c) I-T curves for speaking the words ‘MXene’, ‘Carbon’ and ‘OK’.

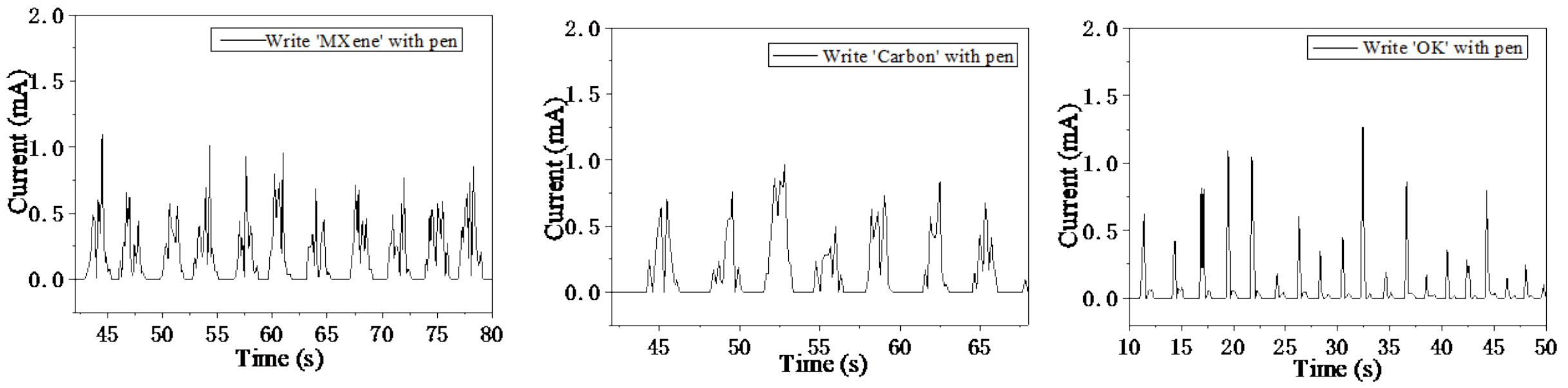

**Figure S10.** (a-c) The I-T curve of writing ‘MXene’, ‘Carbon’ and ‘OK’ with a pen on the sensor.

**Table S1.** Comparison of sensor performance based on MXene material.

| Sensing mechanism | Materials | Detection limit | Sensing range | Sensitivity | response time | Cycling stability | Detection Voltage /power consumption | References |
|---|---|---|---|---|---|---|---|---|
| Piezoresistive | MXene annosheets | 10.2 Pa | 23 Pa – 30 kPa | 0.55 $kPa^{-1}$ (23 to 982 Pa), 3.81 $kPa^{-1}$ (982 to 10 kPa), 2.52 $kPa^{-1}$ (10 to 30 kPa). | 11 ms | 10000 | 0.01V/ $10^{-8}$ W | [26] |
| Piezoresistive | MXene/CS/PU | 9 Pa | 9 Pa – 245.7 kPa | 0.014 $kPa^{-1}$ (<6.5 kPa), 0.015 $kPa^{-1}$ (6.5-85.1 kPa), 0.001 $kPa^{-1}$ (>85.1 kPa). | 19 ms | 5000 | / | [27] |
| Piezoresistive | MXene–textile | / | 5 kPa – 40 kPa | 0.029 $kPa^{-1}$ (5-25 kPa), 3.844 $kPa^{-1}$(25-29 kPa), 12.095 $kPa^{-1}$(29-40 kPa) | 26 ms /52 ms | 5600 | / | [28] |
| Piezoresistive | MXene/rGO | 10 Pa | 0.2 kPa – 3.5 kPa | 4.05 $kPa^{-1}$ (0.2-0.4 kPa), 22.56 $kPa^{-1}$ (1.25-3.4 kPa). | 200 ms | 10000 | / | [30] |
| Piezoresistive | MXene /NMC | 8 Pa | 10 − 200 Pa | 24.63 $kPa^{-1}$ (10−200 Pa) | 14 ms | 5000 | / | [47] |
| Piezoresistive | MXene/PVB | 6.8 Pa | 31.2 Pa – 2.205 MPa | 11.9 $kpa^{-1}$ (31.2Pa -312 Pa), 1.15 $kPa^{-1}$ (312 Pa-62.4 kPa), 0.20 $kPa^{-1}$ (62.4 kPa-1248.4 kPa). | 110 ms | | 0.1mV/ $3.6 \times 10^{-10}$ W | This work |

Movie S1:

The brightness of a LEDs can be controlled by loading different finger pressure on MXene/PVB-based sensor.

Movie S2:

The brightness of a LEDs can be controlled by loading different pressure on MXene/PVB-based sensor with pressure machine.